\documentstyle[12pt,a4wide]{article}
\input epsf

\newcommand{\news}{\setcounter{equation}{0}}
\newcommand{\be}{\begin{equation}}
\newcommand{\ee}{\end{equation}}
\newcommand{\bea}{\begin{eqnarray}}
\newcommand{\eea}{\end{eqnarray}}
\newcommand{\bean}{\begin{eqnarray*}}
\newcommand{\eean}{\end{eqnarray*}}
\font\upright=cmu10 scaled\magstep1
\font\sans=cmss12
\newcommand{\ssf}{\sans}
\newcommand{\stroke}{\vrule height8pt width0.4pt depth-0.1pt}
\newcommand{\Z}{\hbox{\upright\rlap{\ssf Z}\kern 2.7pt {\ssf Z}}}

\newcommand{\C}{{\rlap{\rlap{C}\kern 3.8pt\stroke}\phantom{C}}}
\newcommand{\R}{\hbox{\upright\rlap{I}\kern 1.7pt R}}
\newcommand{\CP}{\C{\upright\rlap{I}\kern 1.7pt P}}
\newcommand{\half}{\frac{1}{2}}
\newcommand{\mt}{\rlap{\ssf T}\kern 3.0pt{\ssf T}}
\newcommand{\spc}{spectral curve }
\newcommand{\ode}{{\scriptsize ODE}}
\newcommand{\ivp}{{\scriptsize IVP}}
\newcommand{\identity}{{\upright\rlap{1}\kern 2.0pt 1}}

\newcommand{\alim}{3^{-5/4}\sqrt{2}}
\newcommand{\sech}{\mbox{\,sech\,}}
\begin{document}
\pagestyle{plain}

\title{\vskip -70pt
\begin{flushright}
{\normalsize DAMTP 95-13} \\
{\normalsize To appear in Communications in Mathematical
Physics} \\
\end{flushright}
\vskip 20pt
{\bf Tetrahedral and Cubic Monopoles} \vskip 10pt}

\author{Conor J. Houghton \\[10pt]
and \\[10pt]
Paul M. Sutcliffe\thanks{
Address from September 1995,
 Institute of Mathematics,
University of Kent at Canterbury, Canterbury CT2 7NZ.
 Email P.M.Sutcliffe@ukc.ac.uk
} \\[20pt]
{\sl Department of Applied Mathematics and Theoretical Physics} \\[5pt]
{\sl University of Cambridge} \\[5pt]
{\sl Silver St., Cambridge CB3 9EW, England}\\[5pt]
{\normalsize c.j.houghton@damtp.cam.ac.uk \& p.m.sutcliffe@damtp.cam.ac.uk}
 \\[10pt]}

\date{March 1995}
\maketitle
\begin{abstract}
Using a numerical implementation of the ADHMN construction, we compute
the fields and energy densities of a charge three monopole with 
tetrahedral symmetry and a charge four monopole with octahedral
symmetry.
 We then construct
a one parameter family of spectral curves and Nahm data which
represent charge four monopoles with tetrahedral symmetry, which
includes the monopole with octahedral symmetry as a special case.
 In the moduli space
approximation, this family describes a novel kind of four
monopole scattering and we use
our numerical scheme to construct the 
energy density at various times during the motion.

\end{abstract}
\newpage
\section{Introduction}
\news
\ \indent BPS monopoles are topological solitons in a Yang-Mills-Higgs gauge
theory in three space dimensions. The equation for static monopoles
is integrable, so that a variety of techniques are available for
studying monopoles and constructing solutions. Monopoles of charge
one and two are well-understood, with explicit solutions known, but 
for higher charges the situation is not so clear. Despite the 
integrability of the equation, explicit solutions for charge three
and above are known only in the axisymmetric case, which corresponds to
coincident monopoles. Very recently, some progress has been made
in this area \cite{HMM} with existence proofs for a charge three
monopole with tetrahedral symmetry and a charge four monopole with
octahedral symmetry. In this paper, we compute these monopoles
using a numerical implementation of the 
Atiyah-Drinfeld-Hitchin-Manin-Nahm
(ADHMN) construction and
display their energy densities.

When time dependence is introduced, the monopole equation of
motion is not integrable. However, analytical progress can still be
made, via the moduli space approximation \cite{M}, from knowledge of
the static monopoles. This has been extensively studied for
the case of charge two monopole scattering \cite{AH}, but the
extension to higher charges has proved a less tractable problem.
We have made some progress in this area and present our results
on a particularly symmetric example of charge four monopole scattering.
The charge four monopole has tetrahedral symmetry throughout the
motion, which is the key to our construction of the relevant
spectral curves and Nahm data. We use our numerical scheme to
display the energy density at various times.

\section{Monopoles, spectral curves and Nahm data}
\news
\ \indent In this paper, we study solutions of the Bogomolny equation
\be
D_i\Phi=-\half\epsilon_{ijk}F_{jk}
\label{bog}
\ee
for SU(2) BPS monopoles in \R$^3$. Here $D_i=\frac{\partial}
{\partial x_i}+[A_i,$ \ is the covariant derivative with $A_i$ the 
{\sl su(2)}-valued gauge potential and $F_{jk}$ the gauge field.
$\Phi$ is the Higgs field, which is an {\sl su(2)}-valued scalar
field satisfying the boundary condition
\be
\|\Phi\|=1-\frac{k}{r}+O(\frac{1}{r^2}) \hskip 10pt\mbox{as}
\hskip 10pt r\rightarrow\infty
\label{bc}
\ee
where $r=\vert\mbox{\boldmath $x$}\vert$, 
$\|\Phi\|^2=-\half\mbox{tr}\Phi^2$
and $k$ is a positive integer, known as the magnetic charge.
We shall refer to a monopole with magnetic charge $k$ as a
$k$-monopole.
The energy density, ${\cal E}$, of a monopole is given by
\be
{\cal E}=-\half\mbox{tr}(D_i\Phi)(D_i\Phi)-\frac{1}{4}
\mbox{tr}(F_{ij}F_{ij}).
\label{energy}
\ee
The energy is the integral of ${\cal E}$ over all space and is
equal to $8\pi k$.

Equation (\ref{bog}) may be obtained by dimensional reduction of the
self-dual Yang-Mills equation in \R$^4$, for which there is a
well-known twistor correspondence; namely that solutions of the
self-duality equations correspond to certain holomorphic vector
bundles over the standard complex 3-dimensional twistor space.
This correspondence may be reduced 
\cite{W,HA,HB} to give that monopoles correspond
to particular holomorphic vector
bundles over a mini-twistor space \mt, which is a 2-dimensional
complex manifold isomorphic to the holomorphic tangent bundle to the
Riemann sphere {\sl ie} \mt$\cong$T\CP$^1$. Moreover, the bundle
(and hence the monopole) is determined by an algebraic curve
in \mt, called the spectral curve, which must satisfy certain 
reality and non-singularity conditions.

The space \mt\ is a fibre bundle over \CP$^1$ with each fibre being a
copy of \C. Let $\zeta$ be the standard coordinate on the base space
and $\eta$ the fibre coordinate, then the three spectral curves
of interest in this paper are
\bea
&\eta^3&+i
\frac{\Gamma(1/6)^3\Gamma(1/3)^3}
{48\sqrt{3}\pi^{3/2}}
\zeta(\zeta^4-1)=0 \label{sca}\\[10pt]
&\eta^4&+
\frac{3\Gamma(1/4)^8}{64\pi^2}
(\zeta^8+14\zeta^4+1)=0 \label{scb}\\[10pt]
&\eta^4&+ i36a\kappa^3\eta\zeta(\zeta^4-1)+
3\kappa^4(\zeta^8+14\zeta^4+1)=0.\label{scc}
\eea
In \cite{HMM} it is proved that (\ref{sca}) is the spectral curve
of a 3-monopole with tetrahedral symmetry and that (\ref{scb})
is the spectral curve
of a 4-monopole with octahedral symmetry. It is the monopole fields
and energy densities which correspond to these two spectral
curves that we shall compute numerically in Section 3. In Section
4 we shall prove that (\ref{scc}) is the spectral curve of a
4-monopole with tetrahedral symmetry for all
$a\in(-\alim,\alim)$ where
$2\kappa$ is the real period of the elliptic curve
\be
y^2=4(x^3-x+3a^2).
\ee
If $a=0$ then (\ref{scc}) becomes (\ref{scb}), so that at this point
the 4-monopole has octahedral symmetry.

Although the \spc approach to monopoles is a very useful and 
powerful technique, its main drawback is that the non-singularity 
constraint, which an algebraic curve must satisfy to be the spectral
curve of a monopole, is rather formidable to check. However, there is
an alternative approach to the construction of monopoles which
nicely complements the \spc formulation, in the sense that the 
non-singularity of the monopole is automatic. The ADHMN construction
\cite{N,HB} is an equivalence between $k$-monopoles and Nahm data
$(T_1,T_2,T_3)$, which are three $k\times k$ matrices which depend
on a real parameter $s\in[0,2]$ and satisfy the following;\\

\newcounter{con}
\setcounter{con}{1}
(\roman{con})  Nahm's equation
\be
\frac{dT_i}{ds}=\half\epsilon_{ijk}[T_j,T_k] \nonumber
\ee\\

\addtocounter{con}{1}
(\roman{con}) $T_i(s)$ is regular for $s\in(0,2)$ and has simple
poles at $s=0$ and $s=2$,\\

\addtocounter{con}{1}
(\roman{con}) the matrix residues of $(T_1,T_2,T_3)$ at each
pole form the irreducible $k$-dimensional representation of SU(2),\\

\addtocounter{con}{1}
(\roman{con}) $T_i(s)=-T_i^\dagger(s)$,\\

\addtocounter{con}{1}
(\roman{con}) $T_i(s)=T_i^t(2-s)$.\\

\setcounter{con}{1}
Equation (\roman{con}) is equivalent to a Lax pair and hence there
is an associated algebraic curve, which is in fact the spectral curve.
Explicitly, the \spc may be read off from the Nahm data as the
equation
\be
\mbox{det}(\eta+(T_1+iT_2)-2iT_3\zeta+(T_1-iT_2)\zeta^2)=0.
\label{curve}
\ee

In Section 3, we review how to obtain the monopole fields from the
Nahm data and explain our numerical implementation of this procedure.

\section{Numerical ADHMN construction}
\news
\ \indent
Finding the Nahm data effectively solves the nonlinear part of the
monopole construction but in order to calculate the fields themselves
the linear part of the ADHMN construction must also be implemented
\cite{N,HB}. Given
Nahm data $(T_1,T_2,T_3)$ for a $k$-monopole we must solve the 
ordinary differential equation (\ode) 
\be
({\identity}_{2k}\frac{d}{ds}+{\identity}_k\otimes x_j\sigma_j
+iT_j\otimes\sigma_j){\bf v}=0
\label{lin}
\ee
for the complex $2k$-vector ${\bf v}(s)$, where $\identity_k$ denotes
the $k\times k$ identity matrix, $\sigma_j$ are the Pauli matrices and
${\bf x}=(x_1,x_2,x_3)$ is the point in space at which the monopole
fields are to be calculated. Introducing the inner product
\be
\langle{\bf v}_1,{\bf v}_2\rangle =\int_0^2 {\bf v}_1^\dagger{\bf v}_2\ ds
\label{ip}
\ee
then the solutions of (\ref{lin}) which we require are those which are
normalizable with respect to (\ref{ip}). It can be shown that the
space of normalizable solutions to (\ref{lin}) has (complex) dimension
2. If $\widehat {\bf v}_1,\widehat {\bf v}_2$ is an orthonormal basis
for this space then the Higgs field $\Phi$ is given by
\be
\Phi=i\left[ \begin{array}{cc}
\langle(s-1)\widehat {\bf v}_1,\widehat {\bf v}_1\rangle &
\langle(s-1)\widehat {\bf v}_1,\widehat {\bf v}_2\rangle \\
\langle(s-1)\widehat {\bf v}_2,\widehat {\bf v}_1\rangle &
\langle(s-1)\widehat {\bf v}_2,\widehat {\bf v}_2\rangle 
\end{array}
\right].
\label{higgs}
\ee
There is a similar expression for the gauge potential but we shall not
need this here. 
In some cases this procedure, which goes from Nahm data to the
Higgs field, may be completed analytically to give an explicit closed
form for $\Phi$. However, the Nahm data which we consider in this
paper is sufficiently complicated that to calculate a closed form
expression for $\Phi$ appears not to be a tractable problem. We
therefore turn to a numerical implementation of the above procedure,
which we now describe.

The first issue we confront in a numerical approach is to calculate
numerical values for the Nahm data on the interval $s\in[0,2]$.
Although we shall have explicit expressions for the Nahm data this is
still not quite a trivial issue, since the expressions involve the
Weierstrass elliptic function $\wp$ and its derivative.
However, we can keep the number of calculations of $T_i(s)$ to a
minimum by noting that if a fixed step \ode\ solver is used to 
integrate (\ref{lin}) then the Nahm data is required at the same $s$ 
values for every integration of (\ref{lin}) for all
initial conditions and ${\bf x}$ positions. Therefore we compute, 
once and for all, $T_i(s)$ at $2P$ equidistant points for $s\in[0,2]$
and store these values, which are then used as a look-up table when 
integrating (\ref{lin}) by a fourth order Runge-Kutta method with
fixed steplength $ds=2P^{-1}$. The values in the look-up table are
computed from the closed form expressions using
{\scriptsize MATHEMATICA}.

Let $\Omega(I)$ denote the space of solutions to (\ref{lin})
which are normalizable for $s$ in the interval $I$. Then we 
require a basis for the 2-dimensional space $\Omega([0,2])$.
The question we now address is how to obtain this basis from
solutions of the initial value problem (\ivp) associated with the
\ode\ (\ref{lin}). Consider the \ivp\ of (\ref{lin}) at the pole
$s=0$, which has the form
\be
s\frac{d{\bf v}}{ds}=B_s{\bf v}
\label{pbc}
\ee
where $B_s$ is a regular $2k\times 2k$ matrix function of 
$s\in[0,2)$. This is a regular-singular problem so that
$\Omega([0,2))$ has dimension $N$, where $N$ is the number of 
positive eigenvalues (counted with multiplicity) of $B_0$.
If $N$ was equal to 2 then we could easily compute a basis for
$\Omega([0,2])$ since it would (almost) be given by a basis for 
$\Omega([0,2))$, which can be found by integrating (\ref{lin}),
as described above, with two different initial conditions.
However, for all the cases considered in this paper we find $N>2$,
so that the problem requires a little more work. By symmetry of the
Nahm data, if we consider the \ivp\ of (\ref{lin}) at the pole $s=2$
(with $ds<0$) then we have a similar regular-singular problem
involving a matrix which again has $N$ positive eigenvalues. By
integrating this \ivp\ we can compute a basis for the $N$-dimensional
space $\Omega((0,2])$. The 2-dimensional space we require is the 
intersection of the above two $N$-dimensional spaces {\sl ie}
\be
\Omega([0,2])=\Omega([0,2))\cap\Omega((0,2]).
\ee
To find the intersection of these two spaces is a shooting problem
but because the \ode\ (\ref{lin}) is linear this shooting problem can
be reduced to linear algebra as follows. 
Let ${\bf u}_1(s),{\bf u}_2(s),..,{\bf u}_N(s)$ denote $N$ 
$2k$-vectors which form a basis for $\Omega([0,2))$ and

\noindent
${\bf u}_{N+1}(s),{\bf u}_{N+2}(s),..,{\bf u}_{2N}(s)$
 a basis
for $\Omega((0,2])$. Explicitly these vectors are computed by
solving the \ivp\ at $s=0$ and $s=2$ with $N$ different initial
conditions each. Define the $2k\times2N$ matrix
\be
U=\left[ 
\begin{array}{cccc}
|&|& &|\\
|&|& &|\\
{\bf u}_1(1)& {\bf u}_2(1)& ...& {\bf u}_{2N}(1) \\
|&|& &|\\
|&|& &|
\end{array}\right]
\label{umatrix}
\ee
then we need to find a basis for the 2-dimensional kernel of $U$
{\sl ie} to solve the matrix equation
\be
U{\bf w}={\bf 0}
\label{keru}
\ee
for ${\bf w}=(w_1,...,w_{2N})\neq{\bf 0}$.
Numerically this is performed by row reduction of the matrix $U$
followed by back substitution. Let ${\bf w}^{(1)}$ and ${\bf w}^{(2)}$ 
denote two independent solutions to (\ref{keru}), then a basis
for $\Omega([0,2])$ is given by
\be
{\bf v}_l(s)=\left\{
\begin{array}{ll}
\ \sum_{j=1}^N {\bf w}_j^{(l)}{\bf u}_j(s) & \mbox{if $0\leq s\leq
1$}\\
 & \\
-\sum_{j=N+1}^{2N} {\bf w}_j^{(l)}{\bf u}_j(s) & \mbox{if $1< s\leq
2$}
\end{array}
\right.
\ee
for $l=1,2$.
To summarize, the above procedure consists in integrating (\ref{lin}) 
$N$ times from each end of the interval $[0,2]$ to the centre
and then finding linear combinations of these solutions such that these
combinations, which start at each end of the interval, match at the centre.

Given ${\bf v}_1$ and ${\bf v}_2$ we use the Gram-Schmidt 
orthonormalization algorithm, with inner product (\ref{ip}) (and
the integral calculated from the $P$ data values via a simple Simpsons
rule), to obtain two orthonormal vectors $\widehat{\bf v}_1,\widehat
{\bf v}_2$. The Higgs field $\Phi$ is then computed according to 
(\ref{higgs}) and to calculate the energy density we make use
of the formula
\be
{\cal E}=\bigtriangleup \|\Phi\|^2
\label{lap}
\ee
where $\bigtriangleup$ denotes the laplacian on \R$^3$.
Numerically we use the above scheme to calculate $\|\Phi\|^2$
on a spatial lattice of $M\times M\times M$ points and approximate
the laplacian in (\ref{lap}) using a finite difference method with
a 7-point stencil. This completes our numerical ADHMN algorithm.

Although every  stage of our algorithm is a relatively 
inexpensive computing task each must be executed many times to build
up a detailed picture of the energy density. To produce each of the
energy density plots appearing later in the paper we used the values
$P=50$ and $M=31$, with
$(x_1,x_2,x_3)\in[-5,5]\times[-5,5]\times[-5,5]$.
This means that the \ode\ (\ref{lin}) must be
solved to the order of $10^5$ times to produce one energy density plot.
Implementing our scheme on a workstation gave a run time of approximately
30 minutes to produce the data for each plot.

The Nahm data which correspond to the spectral curves (\ref{sca})
and (\ref{scb}),
of a 3-monopole with tetrahedral symmetry and a 4-monopole 
with octahedral symmetry respectively, is given in \cite{HMM} 
and we shall make
use of it now. The method used to obtain these data is 
reviewed in Section 4 when we shall use it to calculate the Nahm data
for the spectral curve (\ref{scc}). Not all the  Nahm data given in
 \cite{HMM} explicitly satisfies conditions
\setcounter{con}{4}
(\roman{con})
\addtocounter{con}{1}
and (\roman{con}) given earlier. However the properties of the associated
spectral curves implies that there exists a constant 
$k\times k$ matrix in each case such that conjugation of the Nahm data
by this matrix produces equivalent Nahm data which does satisfy the
conditions, and this is enough. Conjugation by a matrix is equivalent
to a change of basis for the $k$-dimensional representation of SU(2)
formed by the matrix residues of $(T_1,T_2,T_3)$ at the $s=0$ pole.
For our purposes it is convenient if this is a real representation
and so (if necessary) we make a transformation to achieve this.
 In the case $k=3$ the Nahm data from \cite{HMM} is equivalent to
\be
T_1=\left[\begin{array}{ccc}
0&0&0\\
0&0&-z\\
0&\bar z&0
\end{array}
\right]
\;\;
T_2=\left[\begin{array}{ccc}
0&0&-\bar z\\
0&0&0\\
z&0&0
\end{array}
\right]
\;\;
T_3=\left[\begin{array}{ccc}
0&z&0\\
-\bar z&0&0\\
0&0&0
\end{array}
\right]
\ee
where 
\be
z=\frac{\omega\wp^\prime(\omega s)}
{2\wp(\omega s)}+\frac{\sqrt{3}\omega}{\wp(\omega s)}
\mbox{ , }\;\;
\omega=e^{i\pi/6}\frac{\Gamma(1/6)\Gamma(1/3)}
{12\sqrt{\pi}}
\ee
and 
$\wp$ is the Weierstrass function satisfying
\be
\wp^{\prime2}=4\wp^3-4
\ee
where $^\prime$ denotes differentiation with respect to the argument.

With this Nahm data equation (\ref{lin}) is equivalent to the set
of coupled \ode's
\bea
& &\dot v_1+x_3v_1+(x_1+ix_2)v_2+izv_3-\bar zv_6=0
\nonumber\\
& &\dot v_2-x_3v_2+(x_1-ix_2)v_1-izv_4+\bar zv_5=0
\nonumber\\
& &\dot v_3+x_3v_3+(x_1+ix_2)v_4-i\bar zv_1-i zv_6=0\\
& &\dot v_4-x_3v_4+(x_1-ix_2)v_3+i\bar zv_2-i zv_5=0
\nonumber\\
& &\dot v_5+x_3v_5+(x_1+ix_2)v_6+zv_2+i\bar zv_4=0
\nonumber\\
& &\dot v_6-x_3v_6+(x_1-ix_2)v_5-zv_1+i\bar zv_3=0
\nonumber
\eea

where $\dot v_1=\frac{dv_1}{ds}$ etc.
In terms of the notation introduced earlier we find that the 
matrix $B_0$ has eigenvalues $\{1,1,1,1,-2,-2\}$ so that $N=4$.

Figure 1. displays the output of our algorithm for this case. The
plot shows a surface of constant energy density ${\cal E}=0.20$.
The tetrahedral symmetry of this surface is clearly evident and
plots for other values of ${\cal E}$ close to this one
are qualitatively similar.
For large values of ${\cal E}$ the surface breaks up into four
disconnected pieces centered on the vertices of a regular
tetrahedron.

% put FIG 1 here
\begin{figure}[ht]
\begin{center}
\leavevmode
\epsfxsize=10cm \epsffile{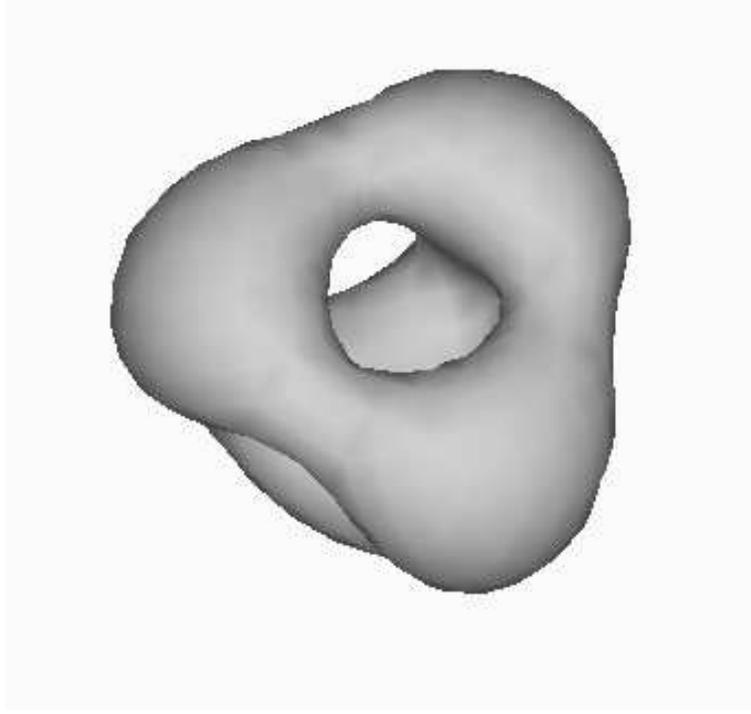}
\caption{Tetrahedral 3-monopole; surface of constant energy density
${\cal E}=0.20$.}
\end{center}
\end{figure}

We now turn to the 4-monopole with octahedral symmetry.
The Nahm data from
\cite{HMM} (after a change of basis) is
$$
T_1=i\left[\begin{array}{cccc}
0&\sqrt{3}(2y-x)&0&-10y\\
\sqrt{3}(2y-x)&0&-6y-2x&0\\
0&-6y-2x&0&\sqrt{3}(2y-x)\\
-10y&0&\sqrt{3}(2y-x)&0
\end{array}
\right]
$$
$$
T_2=\left[\begin{array}{cccc}
0&\sqrt{3}(2y-x)&0&10y\\
-\sqrt{3}(2y-x)&0&-6y-2x&0\\
0&6y+2x&0&\sqrt{3}(2y-x)\\
-10y&0&-\sqrt{3}(2y-x)&0
\end{array}
\right]
$$
\be
T_3=i\left[\begin{array}{cccc}
-4y-3x&0&0&0\\
0&12y-x&0&0\\
0&0&-12y+x&0\\
0&0&0&4y+3x
\end{array}
\right]
\ee
where
\be
x=\omega e^{i\pi/4}\frac{(5\wp^2(s\omega
e^{i\pi/4}/2)-3)}{10\wp^\prime(s\omega
e^{i\pi/4}/2)}
, \
y=\omega e^{i\pi/4}\frac{1}{10\wp^\prime(s\omega
e^{i\pi/4}/2)}
, \
\omega=\frac{\Gamma(1/4)^2}{\sqrt{8\pi}}
\ee
and 
$\wp$ is the Weierstrass function satisfying
\be
\wp^{\prime2}=4\wp^3-4\wp.
\ee 
Then equation (\ref{lin}) is equivalent to 
\bea
& &\dot v_1+x_3v_1+(x_1+ix_2)v_2
+(4y+3x)v_1+20yv_8=0
\nonumber\\
& &\dot v_2-x_3v_2+(x_1-ix_2)v_1
-(4y+3x)v_2+2\sqrt{3}(-2y+x)v_3=0
\nonumber\\
& &\dot v_3+x_3v_3+(x_1+ix_2)v_4
+2\sqrt{3}(-2y+x)v_2+(-12y+x)v_3=0
\nonumber\\
& &\dot v_4-x_3v_4+(x_1-ix_2)v_3
+(12y-x)v_4+4(3y+x)v_5=0
\nonumber\\
& &\dot v_5+x_3v_5+(x_1+ix_2)v_6
+4(3y+x)v_4+(12y-x)v_5=0
\nonumber\\
& &\dot v_6-x_3v_6+(x_1-ix_2)v_5
+(-12y+x)v_6+2\sqrt{3}(-2y+x)v_7=0
\nonumber\\
& &\dot v_7+x_3v_7+(x_1+ix_2)v_8
+2\sqrt{3}(-2y+x)v_6-(4y+3x)v_7=0
\nonumber\\
& &\dot v_8-x_3v_8+(x_1-ix_2)v_7
+20yv_1+(4y+3x)v_8=0.
\nonumber\\
\eea
We find that the matrix $B_0$ has eigenvalues
$\{\frac{3}{2},\frac{3}{2},\frac{3}{2},
\frac{3}{2},\frac{3}{2},-\frac{5}{2},
-\frac{5}{2},-\frac{5}{2}\}$, so that $N=5$.

Figure 2. displays the output of our algorithm in this case. 
The
plot shows a surface of constant energy density ${\cal E}=0.14$.
Note that for the monopole with octahedral symmetry
a constant energy density
surface could have resembled an octahedron or a cube;
clearly it is the latter. 
It is therefore more natural to refer to this monopole not
as an octahedral monopole but as a cubic monopole.
For large values of ${\cal E}$
the surface breaks up into eight disconnected pieces on the
vertices of a cube.

% Put FIG 2. here
\begin{figure}[ht]
\begin{center}
\leavevmode
\epsfxsize=10cm\epsffile{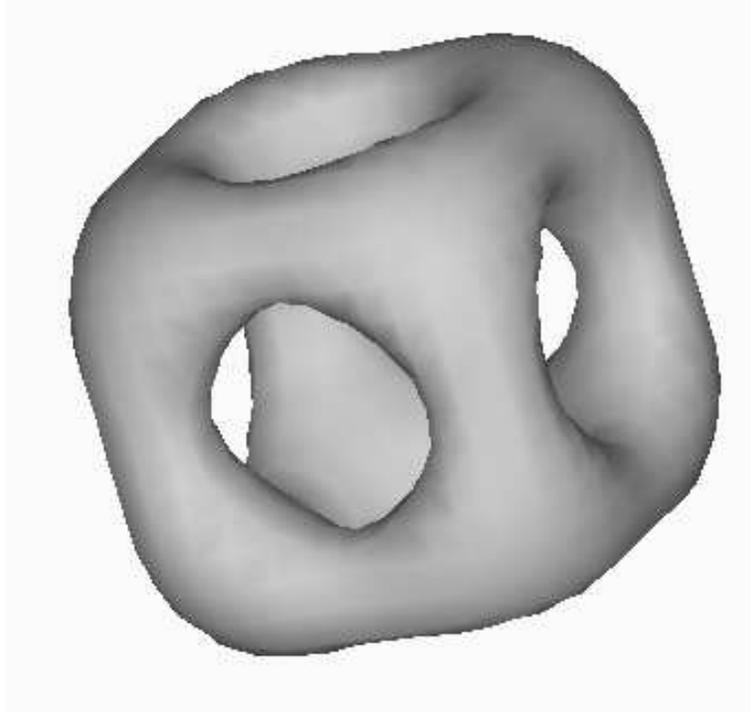}
\caption{Cubic 4-monopole; surface of constant energy density
${\cal E}=0.14$.}
\end{center}
\end{figure}

\section{Four Monopole Scattering}
\news\ \indent
In this Section we shall 
use the method of \cite{HMM} to
construct a one parameter family of Nahm data,
which represent four monopoles with tetrahedral symmetry.
The imposition of tetrahedral symmetry facilitates the solving of 
Nahm's equations, since we will only have to consider Nahm data which are
invariant under the action of the tetrahedral group $T\subset SO(3)$. 

The Nahm data are an \R$^3\otimes sl(k,\C)$ valued function of $s$, which
transform under the rotation group $SO(3)$ as 
\be\underline{3}\otimes
sl(\underline{k})\label{translaw}\ee
where $\underline{r}$ denotes the unique irreducible r dimensional
representation of $su(2)$. Since $gl(\underline{k})\cong
\underline{k}\otimes\underline{k}$ Clebsh-Gordon decomposition gives
$$ gl(\underline{k})\cong\underline{2k-1}\oplus
\underline{2k-3}\oplus \ldots \oplus \underline{3}\oplus\underline{1}
$$
and so
\be sl(\underline{k})\cong\underline{2k-1}\oplus
\underline{2k-3}\oplus \ldots \oplus \underline{3}
\label{decomp}
\ee
Substituting into (\ref{translaw})
\begin{eqnarray}\underline{3}\otimes sl(\underline{k}) &\cong&\underline{3}\otimes (\underline{2k-1}\oplus\underline{2k-3}\oplus \ldots \oplus \underline{3})\nonumber\\
&\cong&( \underline{2k+1}\oplus \underline{2k-1}\oplus
\underline{2k-3})\oplus \ldots \oplus (
\underline{5}\oplus \underline{3}\oplus\underline{1}).
\end{eqnarray}
Thus the Nahm data corresponding to four monopoles are in the carrier
space $( \underline{9}\oplus \underline{7}\oplus
\underline{5})\oplus (\underline{7}\oplus \underline{5}\oplus
\underline{3}) \oplus (
\underline{5}\oplus \underline{3}\oplus\underline{1})$.   The
$\underline{1}$ representation is, of course, invariant under all of
$SO(3)$. We will calculate this and the other tetrahedral invariants.
 
Write $X,Y$ and $H$ for the basis of $su(2)$ satisfying the
commutation relations
\be [X,Y]=H;\qquad [H,X]=2X;\qquad [H,Y]=-2Y.\ee
These may be represented by the principal 
 $su(2)$ subalgebra of $sl(\underline{k})$ which in
turn acts on the algebra by the adjoint action. In this
representation  $X$ is a rank $k-1$
nilpotent element and a basis of $sl(\underline{k})$ can be
generated by acting with $Y$ on $X^r$, for $r=1,2,..,k-1$.
Thus
\begin{tabbing}
$X^{k-1}$ \qquad \= $(adY)X^{k-1}$ \qquad \= $(adY)^2 X^{k-1}$ \qquad
\= \ldots \qquad \= \ldots \qquad \= $(adY)^{2k-2} X^{k-1}$\\
\vdots\>\ldots\\
$X^r$\> $(adY)X^r$\>$(adY)^2 X^r$\> \ldots\>$(adY)^{2r} X^r$\\
\vdots\>\ldots\\
$X$ \qquad\ \>$(adY)X$\>$(adY)^2 X$
\end{tabbing}
is a basis of $sl(\underline{k})$. The element $X^r$ of the abelian
nilpotent subalgebra $\langle X,X^2,\ldots\,X^{k-1}\rangle$ is the
highest weight vector for the $su(2)$
representation $\underline{2r+1}$ lying in the decomposition
(\ref{decomp}) of $sl(\underline{k})$.  

It is convenient to exploit the representations of $su(2)$ on 
homogeneous polynomials over \CP$^1$
since the invariant homogeneous polynomials are known 
\cite{K}, and also it connects with the spectral curve approach.
The $r+1$ dimensional $su(2)$ representation $\underline{r+1}$ is defined
on degree $r$ homogeneous polynomials by the identification 
\be X=\zeta_1 \frac{\partial}{\partial \zeta_0};\qquad Y=\zeta_0
\frac{\partial}{\partial \zeta_1};\qquad H=-\zeta_0
\frac{\partial}{\partial \zeta_0}+\zeta_1 \frac{\partial}{\partial
\zeta_1}.\ee
In the case of degree $r$ homogeneous polynomials we can identify
highest weight vector $\zeta_1^{r}$ and basis $\{\zeta_1^r,(\zeta_0\frac{\partial}{\partial\zeta_1})\zeta_1^r,\/\ldots\/,(\zeta_0\frac{\partial}{\partial\zeta_1})^r\zeta_1^r)\}$. Thus we can relate a degree $2r$ homogeneous polynomial $q_{2r}(\zeta_0,\zeta_1)$ and a matrix $S$ in the $\underline{2r+1}$ representation of the decomposition of
$sl(\underline{k})$ by rewriting
$q_{2r}(\zeta_0,\zeta_1)$ as $q_{2r}(\zeta_0\frac{\partial}{\partial\zeta_1})\zeta_1^{2r}$
and then letting  
\be S=q_{2r}(adY)X^{r}\label{constr}.\ee

The lowest degree $T$-invariant polynomial is of degree $6$. It is 
\be\zeta_1^5 \zeta_0-\zeta_1\zeta_0^5.\label{tet}\ee
There is a degree $8$ $T$-invariant polynomial
\be\zeta_1^8+14\zeta_1^4\zeta_0^4+\zeta_0^8\label{cub}\ee
which is also invariant under the octahedral group.   Thus, in addition
to the $SO(3)$ invariant there are $T$-invariant Nahm triplets lying
in the $\underline{9}$ representation and in both the $\underline{7}$
representations. It is convenient to write $( \underline{9}_u\oplus \underline{7}_m\oplus
\underline{5}_l)\oplus (\underline{7}_u\oplus \underline{5}_m\oplus
\underline{3}_l) \oplus (
\underline{5}_u\oplus \underline{3}_m\oplus\underline{1}_l)$ so that
we may distinguish $\underline{7}_m$ and $\underline{7}_u$.

We can now construct the $T$-invariant Nahm triplets in
$\underline{9}_u$ and  $\underline{7}_u$ by (\ref{constr}) and the
inclusion
\begin{eqnarray}\underline{2r+1}&\hookrightarrow&\underline{3}\otimes
  \underline{2r-1}\cong\underline{2r+1}\oplus
\underline{2r-1}\oplus\underline{2r-3}\nonumber\\p_{2r}(\zeta_0,\zeta_1)
&\mapsto&\xi_1^2\otimes \frac{\partial^2 p_{2r}}
{\partial\zeta_1^2}+2\xi_0\xi_1\otimes\frac{\partial^2 p_{2r}}
{\partial\zeta_0\partial\zeta_1}+\xi_0^2\otimes 
\frac{\partial^2 p_{2r}}{\partial\zeta_0^2}.\end{eqnarray}
We choose matrices  
\be X=\left[ \begin{array}{rcrc}0 & \sqrt {3} & 0 & 0 \\0 & 0 & 2 & 0
\\0 & 0 & 0 & \sqrt {3} \\0 & 0 & 0 & 0\end{array}\right]\;
Y=\left[ \begin{array}{crcr}0 & 0 & 0 & 0 \\ \sqrt {3} & 0 & 0 & 0 \\
0 & 2 & 0 & 0 \\0 & 0 & \sqrt {3} & 0\end{array}\right]\;
\\ H=\left[ \begin{array}{rrrr}3 & 0 & 0 & 0 \\
0 & 1 & 0 & 0 \\0 & 0 & -1 & 0 \\0 & 0 & 0 & -3\end{array}\right] .\ee
Polarizing (\ref{tet}) yields
\be\xi_1^2\otimes (20\zeta_1^3\zeta_0)+2\xi_1\xi_0\otimes 
(5\zeta_1^4-5\zeta_0^4)-\xi_0^2\otimes (20\zeta_1\zeta_0^3)\ee
which we put into the form
\begin{eqnarray}\xi_1^2&\otimes&
  5(\zeta_0\frac{\partial}{\partial
    \zeta_1})\zeta_1^4+(\xi_0\frac{\partial}{\partial\xi_1})\xi_1^2\otimes
 [5-\frac{5}{24}(\zeta_0\frac{\partial}{\partial\zeta_1})^4]\zeta_1^4
\nonumber\\&+&\frac{1}{2}(\xi_0\frac{\partial}{\partial\xi_1})^2\xi_1^2\otimes
 [-\frac{5}{6}(\zeta_0\frac{\partial}{\partial\zeta_1})^3\zeta_1^4]
\end{eqnarray}
and convert to $k\times k$ matrices 
\be X\otimes 
  5adYX^2+adYX\otimes [5-\frac{5}{24}(adY)^4] X^2 +
  \frac{1}{2}(adY)^2 X\otimes [-\frac{5}{6}(adY)^3]X^2\label{adjtet}.\ee
These matrices were calculated explicitly using {\scriptsize
  MAPLE} and are proportional to
\be 
Z_1=\left[ {\begin{array}{rcrc}0 & 2\sqrt{3} & 0 & 0 \\
0 & 0 & 0 & 0 \\0 & 0 & 0 &  -2\sqrt{3} \\0 & 0 & 0 & 0\end{array}} \right]\;
 Z_2=\left[ {\begin{array}{cccc}
0 & 0 & \sqrt{3} & 0 \\0 & 0 & 0 &  \sqrt{3}\\ - \sqrt{3} & 0 & 0 & 0 \\
0 &  -  \sqrt{3} & 0 & 0\end{array}}
 \right] \; 
Z_3=\left[ {\begin{array}{crcr}0 & 0 & 0 & 0 \\ \sqrt{3} & 0 & 0 & 0 \\
0 & 0 & 0 & 0 \\0 & 0 &  -\sqrt{3} & 0\end{array}}\right]
\\
\ee

Similarly polarizing (\ref{cub}) yields
\be\xi_1^2\otimes
(56\zeta_1^6+168\zeta_1^2\zeta_0^4)+2\xi_1\xi_0\otimes
(224\zeta_1^3\zeta_0^3)+\xi_0^2\otimes
(168\zeta_1^4\zeta_0^2+56\zeta_0^6)\ee
which becomes
\begin{eqnarray}\xi_1^2&\otimes&
 [56+\frac{7}{15}(\zeta_0\frac{\partial}{\partial\zeta_1})^4]
\zeta_1^6+(\xi_0\frac{\partial}{\partial\xi_1})\xi_1^2\otimes
  \frac{28}{15}(\zeta_0\frac{\partial}{\partial\zeta_1})^3
  \zeta_1^6 \nonumber\\ &+&\frac{1}{2}
  (\xi_0\frac{\partial}{\partial\xi_1})^2
  \xi_1^2\otimes
[\frac{28}{5}(\zeta_0\frac{\partial}{\partial\zeta_1})^2
+\frac{7}{90}(\zeta_0\frac{\partial}{\partial\zeta_1})^6]\zeta_1^6
\end{eqnarray}
yielding invariant Nahm triplet 
$$ Y_1=\left[ {\begin{array}{crcr}0 & 0 & 0 & -20 \\ 4\,\sqrt {3} &
  0 & 0 & 0 \\ 0 & -12 & 0 & 0 \\ 0 & 0 & 4\,\sqrt {3} &
0\end{array}}\right] 
\; \;
Y_2=\left[ {\begin{array}{rrrr}-4 & 0 & 0 & 0 \\0 & 12 & 0 & 0 \\0 & 0
  & -12 & 0 \\0 & 0 & 0 & 4\end{array}}\right]
\;\;$$
\be
 Y_3=\left[
{\begin{array}{rcrc}0 & -2\,\sqrt {3} & 0 & 0 \\0 & 0 & 6 & 0 \\0 & 0 & 0
  & -2\,\sqrt {3} \\10 & 0 & 0 & 0\end{array}} \right]
\ee  

We calculate the $T$-invariant in $\underline{7}_m$ by constructing an
isomorphism between it and $\underline{7}_u$. We observe that
$X\otimes X^r$ is a highest weight vector of the representation
$\underline{2r+3}_u$ and a basis can be generated by successive
application of $(adY\otimes 1 + 1\otimes adY)$.  Thus, for example, the
invariant (\ref{adjtet}) can be written 
\be [5(adY\otimes 1 + 1\otimes adY)-\frac{1}{24}(adY\otimes 1 +
1\otimes adY)^5]X\otimes X^2.\ee
The highest weight vector for $\underline{2r+3}_m$ is easily calculated
by noting that it is annihilated by $(adX\otimes 1+1\otimes adX)$. It is $(adYX\otimes
X^{r+1}-\frac{1}{r+1}X\otimes adYX^{r+1})$. We can then map
\be\underline{2r+3}_u \stackrel{\cong}{\longrightarrow} \underline{2r+3}_m\nonumber\ee
by
\be  X\otimes X^r\mapsto (adYX\otimes
X^{r+1}-\frac{1}{r+1}X\otimes adYX^{r+1}).\nonumber\ee
Thus the $\underline{7}_m$ invariant is 
\begin{eqnarray}\lefteqn {[5(adY\otimes 1 + 1\otimes adY)-\frac{1}{24}(adY\otimes 1 +
1\otimes adY)^5](adYX\otimes
X^3-\frac{1}{3}X\otimes
adYX^3)}\nonumber\\  &=&X\otimes(-\frac{5}{3}(adY)^2+\frac{1}{72}(adY)^6)X^3+adYX\otimes
(\frac{10}{3}(adY)+\frac{1}{36}(adY)^5)X^3\nonumber\\&\qquad&+(adY)^2X\otimes
(5-\frac{5}{72}(adY)^4)X^3\end{eqnarray}
with corresponding matrices
$$ W_1=\left[ {\begin{array}{rcrc}0 &  - 2\,\sqrt {3} & 0 & 0 \\
0 & 0 & 6 & 0 \\0 & 0 & 0 &  - 2\,\sqrt {3} \\-6 & 0 & 0 &
0\end{array}} \right]
\;\;
 W_2=\left[ {\begin{array}{cccc}0 & 0 & 2\,\sqrt {3} & 0 \\
0 & 0 & 0 &  - 2\,\sqrt {3} \\2\,\sqrt {3} & 0 & 0 & 0 \\
0 &  - 2\,\sqrt {3} & 0 & 0\end{array}} \right]
$$
\be
 W_3=\left[ {\begin{array}{crcr}0 & 0 & 0 & 3 \\
\sqrt {3} & 0 & 0 & 0 \\0 & -3 & 0 & 0 \\0 & 0 & \sqrt {3} & 0
\end{array}}\right] .
\ee

The easiest way of calculating the $SO(3)$ invariant is to observe
that it is annihilated by both $(adX\otimes 1 + 1\otimes adX)$ and
$(adY\otimes 1 +1\otimes adY)$. It is
\be X\otimes (adY)^2X-adYX\otimes adYX+(adY)^2X\otimes X=X
\otimes(-2Y)-adYX\otimes H+(adY)^2X\otimes X\ee

We now change basis so that the $SO(3)$ invariant is given by
$(\rho_1,\rho_2,\rho_3)$, the $su(2)$ basis satisfying
$[\rho_1,\rho_2]=2\rho_3$ etc;
\be \rho_1=X-Y;\qquad \rho_2=i(X+Y);\qquad \rho_3=iH.\nonumber\ee
Thus
\be (Y_1,Y_2,Y_3)\rightarrow
(Y_1^{\prime},Y_2^{\prime},Y_3^{\prime})=(\frac{1}{2}Y_1+Y_3,-\frac{i}{2}
Y_1+iY_3,-iY_2)\ee
and similarly for $(Z_1,Z_2,Z_3)$ and $(W_1,W_2,W_3)$. We drop the
primes on the transformed quantities.

With a view to calculating the Nahm equations the commutation
relations satisfied by the invariant Nahm vectors were calculated
using {\scriptsize MAPLE} 
\be\begin{array}{ll}
[ Y_1,Y_2 ]=-48\rho_3-8Y_3,&
  {[} Z_1,Z_2]=\frac{6}{5}\rho_3+\frac{3}{5}Y_3\\
\\
{[}W_1,W_2]=\frac{12}{5}\rho_3+\frac{6}{5}Y_3,&
{[}\rho_1,Y_2]+[Y_1,\rho_2]=-6Y_3\\
\\
{[}\rho_1,Z_2]+[Z_1,\rho_2]=-4Z_3,&
{[}\rho_1,W_2]+[W_1,\rho_2]=2W_3\\
\\
{[}Y_1,Z_2]+[ Z_1,Y_2]=-32Z_3,&
[ Z_1,W_2]+[ W_1,Z_2]=0\\
\\
{[}Y_1,W_2]+[ W_1,Y_2]=16W_3
\end{array}\nonumber\ee
Writing 
\be T_i(s)=x(s)\rho_i+y(s)Y_i+z(s)Z_i+w(s)W_i\ee
 the Nahm equation $\frac{dT_3}{ds}=[T_1,T_2]$ reduces under the
requirement of $T$-symmetry to the set of coupled nonlinear equations
\begin{eqnarray}\frac{dx}{ds}&=&2x^2-48y^2+\frac{6}{5}z^2+\frac{12}{5}w^2,
   \label{x}
   \\\frac{dy}{ds}&=&-8y^2+\frac{3}{5}z^2+\frac{6}{5}w^2-6xy, 
 \label{ydot} \\\frac{dz}{ds}&=&-4xz-32yz,  \label{zdot}
  \\\frac{dw}{ds}&=&2xw+16wy.  \label{w}  \end{eqnarray}
Calculation of the polynomial
$\mbox{det}(\eta+(T_1+iT_2)-2iT_3\zeta+(T_1-iT_2)\zeta^2)$ 
gives the spectral curve
\be\eta^4+c_1\eta\zeta(\zeta^4-1)+c_1c_2(\zeta^8+14\zeta^4+1)=0\ee
where
\be c_1=288z(x^2+4y^2+3w^2-4xy)\equiv288ic_1^\prime \label{const}\ee
and
\be
c_2=-\frac{48}{288z}(60y^2+3z^2-3w^2+20xy)
\equiv\frac{48}{288}ic_2^\prime
\label{constconst}\ee
are constants.

To solve these equations, we observe that $w$ can be set to zero,
so we do so. We let $u=x-2y$ and $v=x+8y$ to get
\be\frac{du}{ds}=2uv\label{udot}\ee
\be z=i\frac{c_1^\prime}{u^2}\ee
and
\be c_2^\prime = \frac{u^2}{c_1^\prime} \left[ v^2-u^2 -\frac{3c_1^{\prime
    2}}{u^4}\right].\ee
Define $\kappa^4\equiv -16c_1^\prime c_2^\prime$ and $a\equiv8{c_1^{\prime
}}/\kappa^3$ to obtain 
\be4\frac{du}{ds}=-\sqrt{64u^4-4\kappa^4+3a^2\kappa^6u^{-2}}.\ee
Let $t=\kappa s$ and $u(s)=-\kappa\sqrt{\wp(t)}/2$ giving
\be(\frac{\kappa^2}{\sqrt{\wp(t)}}\frac{d\wp(t)}{dt})^2
=4\kappa^4(\wp(t)^2-1+\frac{3a^2}{\wp(t)})\nonumber.\ee
Thus $\wp(t)$ is the Weierstrass function satisfying
\be \wp^{\prime 2}=4\wp^3-4\wp+12a^2.\ee
Hence (\ref{x})-(\ref{w}) are solved by
\begin{eqnarray} x(s)&=&\frac{\kappa}{5}\left(-2\sqrt{\wp(\kappa
    s)}+\frac{1}{4}\frac{\wp^\prime(\kappa s)}{\wp(\kappa
    s)}\right)\\
y(s)&=&\frac{\kappa}{20}\left(\sqrt{\wp(\kappa
    s)}+\frac{1}{2}\frac{\wp^\prime(\kappa s)}{\wp(\kappa s)}\right)\\
z(s)&=&\frac{ia\kappa}{2\wp(\kappa s)}\\w(s)&=&0\end{eqnarray}

In order to determine that these Nahm data correspond to a monopole we
need to examine the boundary conditions. As $t\rightarrow 0$,
$\wp(t)\sim t^{-2}$ and so 
\be x\sim -\frac{1}{2s},\qquad y\sim 0,\qquad z\sim 0.\ee
Therefore at $s=0$ the residue of $T_i$ is $-\frac{1}{2}\rho_i$ and so
it forms an irreducible representation of SU(2). As $t\rightarrow
2\omega_1$ (the real period of the elliptic function $\wp(t)$)
\be x\sim -\frac{3}{10\widetilde s},\qquad y\sim
 \frac{1}{10\widetilde s},\qquad z\sim 0\ee
where $\widetilde s =2\omega_1\kappa^{-1}-s$,
and so the residue of $T_i$ is $R_i=-\frac{3}{10}\rho_i+\frac{1}{10}
Y_i$. The eigenvalues of $2iR_3$ are $\{3,1,-1,-3\}$ demonstrating that
the $R_i$'s are an irreducible representation of SU(2). Furthermore
the functions $x,y$ and $z$ are analytic for $t\in (0,2\omega_1)$. We
set $\kappa=\omega_1$, so that the poles occur at $s=0$ and
$s=2$. This demonstrates the existence of a one parameter
family of monopoles with spectral curves
\be\eta^4+i36a\kappa^3\eta\zeta(\zeta^4-1)+3\kappa^4(\zeta^8+14\zeta^4+1)=0.
\label{monopole} \ee

A single monopole with position $(x_1,x_2,x_3)$ has spectral curve
\be\eta-(x_1+ix_2)+2x_3\zeta+(x_1-ix_2)\zeta^2=0.\ee
The product of four spectral curves corresponding to four monopoles
positioned at the vertices
\be 
\{(+b,+b,+b),(+b,-b,-b),(-b,-b,+b),(-b,+b,-b)\}
\label{tetone}
\ee
 of a regular tetrahedron (where $b>0$) is
\be\eta^4-16ib^3\eta(\zeta^5-\zeta)+4b^4(\zeta^8+14\zeta^4+1)=0.\ee
The spectral curve (\ref{monopole}) has this form when 
\be a=-\alim .\ee
Examination of the integral expression for $\kappa$
\be\kappa = \int_0^X \frac{dx}{\sqrt{1-x^4+3a^2x^6}}\ee
where $X$ is the first positive real root of $0=1-x^4+3a^2x^6$, shows
that $\kappa\rightarrow\infty$ as $a\rightarrow\pm\alim$
but it is finite for $a\in (-\alim,\alim)$.
We conclude that as $a$ approaches $-\alim$ (\ref{monopole})
describes the superposition of four well-separated monopoles 
on the the vertices (\ref{tetone}) of a tetrahedron, with
the distance between monopoles equal to $3^{1/4}\kappa$.
The tetrahedron dual to the one above has vertices
\be 
\{(-b,-b,-b),(-b,+b,+b),(+b,+b,-b),(+b,-b,+b)\}
\ee
with a corresponding product of spectral curves given by
\be\eta^4+16ib^3\eta(\zeta^5-\zeta)+4b^4(\zeta^8+14\zeta^4+1)=0.\ee
Clearly this is the form of the \spc (\ref{monopole}) when $a=\alim$.

If $a=0$ then $z=0$ and $\kappa$ is given by
\be\kappa = \int_0^1 \frac{dx}{\sqrt{1-x^4}}=
\frac{\Gamma(1/4)^2}{\sqrt{8\pi}}
\ee
so that the  spectral curve becomes that of the cubic 
4-monopole given by (\ref{scb}).

We have derived a one parameter family of 4-monopoles with tetrahedral
symmetry. They correspond to a one parameter family of spectral
curves. We can use this one parameter family to discuss low energy
scattering of 4-monopoles because it is a geodesic in the 4-monopole
moduli space. In order to prove that the one parameter family is a geodesic in the 4-monopole moduli space we must allow for the
possibility that the function $w$ is non-zero. We will find that
solutions to the tetrahedral Nahm equations (\ref{x})-(\ref{w})
correspond to the same one parameter family of spectral curves
irrespective of whether or not $w$ is set to zero. This
means that the fixed point set of the tetrahedral symmetry in the
4-monopole moduli space is one dimensional and since the fixed point
set of a group action on the moduli space is totally geodesic this
means that this one parameter family is a geodesic.

If $w$ is not set to zero, we find that
$$\frac{dw}{ds}=2wv$$
and that the equation for $\frac{du}{ds}$, (\ref{udot}) is
unchanged. Thus 
\be \frac{d}{ds}\left(\frac{w}{u}\right)=0\label{divwoveru}\ee
which implies that $w\propto u$. For convenience we choose the
constant to be $\frac{1}{\sqrt{3}}\sinh{\theta}$.
Furthermore if we set 
$$\tilde{u}^2=u^2+3w^2$$
then $\tilde{u}$ satisfies exactly the same equations as were formerly
satisfied by $u$ and replaces $u$ in the expressions for the two
constants, $c_1^{\prime}$ and $c_2^{\prime}$
ie.
\be\frac{d\tilde{u}}{ds}=2\tilde{u}v\label{nudot}\ee
\be z=i\frac{c_1^\prime}{\tilde{u}^2}\ee
and
\be c_2^\prime = \frac{\tilde{u}^2}{c_1^\prime} \left[ v^2-\tilde{u}^2
 -\frac{3c_1^{\prime
    2}}{\tilde{u}^4}\right].\ee

 The solutions $\tilde u,v$ are identical to the earlier solutions $u,v$ 
in the $w=0$ case so that now
\begin{eqnarray}x&=&\frac{4\tilde{u}\sech\theta+v}{5},\\
                y&=&\frac{v-\tilde{u}\sech{\theta}}{10},\nonumber\\
                z&=&i\frac{c_1^{\prime}}{\tilde{u}^2},\nonumber\\
                w&=&\frac{\tilde{u}\tanh{\theta}}{\sqrt{3}}.\end{eqnarray}
It can be seen by explicit calculation that the matrix residues at both ends
of the intervals are irreducible representations irrespective of the
value of $\theta$. Thus the Nahm data always corresponds to a monopole.
However, it is clear from the above construction that
the constants $c_1^{\prime}$ and $c_2^{\prime}$ are independent
of $\theta$. Hence changing the value of $\theta$ does not change
the spectral curve. There is a one-to-one correspondence
(up to gauge transformations) between monopoles and spectral curves,
so changing the value of $\theta$ simply corresponds to a gauge transformation
of the Nahm data. So, by a suitable choice of gauge we can set $\theta=0$
without loss of generality.
This means that we have
arrived at a one parameter family of monopoles by forcing the 4-monopole to admit tetrahedral symmetry. This proves that the family
of monopoles is a geodesic in the 4-monopole moduli space.

In the moduli space approximation \cite{M} the dynamics
of $k$ monopoles is approximated
by geodesic motion on the $k$-monopole moduli space ${\cal M}_k$.
In this Section we have identified a totally
geodesic one-dimensional submanifold of ${\cal M}_4$ and so we can use
the moduli space approximation to convert this into a result on 
four-monopole scattering. Since our submanifold is one-dimensional the
explicit form of the metric is not important. The information we
lose by not knowing the metric is how physical time is related
to the parameter $a$, but this is not serious. The above results,
therefore, have the following interpretation in terms of monopole 
scattering. Four monopoles approach from infinity on the vertices 
of a contracting regular tetrahedron, coalesce to form a configuration
with instantaneous octahedral symmetry, and emerge on the vertices of
an expanding tetrahedron dual to the incoming one. 

% put FIG 3 here
\begin{figure}[ht]
\begin{center}
\leavevmode
\epsfxsize=10cm \epsffile{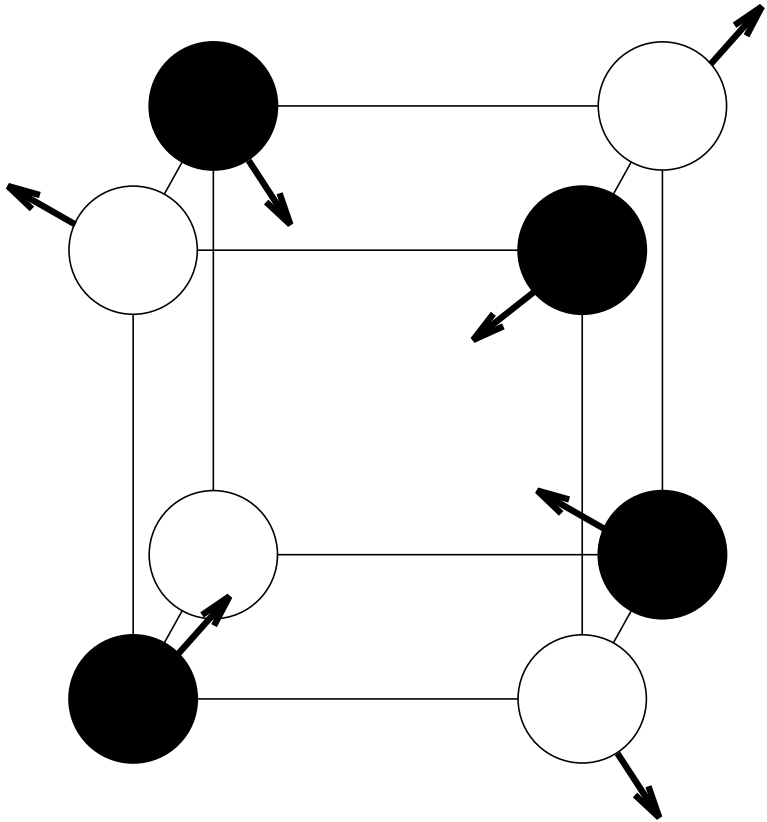}
\caption{Schematic representation of 4-monopole scattering.}
\end{center}
\end{figure}

To make the above scattering process a little clearer we give a
schematic representation in Fig 3. We draw a cube whose centre is
at the origin and whose edges are parallel to the coordinate axes;
it is to be associated with the cubic 4-monopole 
(compare Fig 2.). The incoming monopoles are represented by black
spheres and the outgoing monopoles by white spheres, with an arrow
indicating the direction of motion for each. Note that if one tried
to extend this asymptotic interpretation to the region in which the 
monopoles are close together then one would conclude that the
monopoles suffer no deflection and simply pass through each other.
But this is misleading, since each of the outgoing monopoles
cannot be identified with a single incoming monopole but is a
composition of all the incoming ones. (A similar misleading
interpretation exists for the scattering of three topological
solitons in the plane. For $k$ solitons in the plane with cyclic
$C_k$ symmetry the solitons scatter through an angle $\pi/k$, which
for $k=3$ could mistakenly be taken for zero scattering angle).

To obtain a true picture of the scattering process, one needs to
examine the energy density during the motion. Using our numerical scheme
we can do this. Fig 4.\footnote{Fig 4. is not available in the
hep-th version of this paper.
A hard copy is available on request to P.M.Sutcliffe@ukc.ac.uk,
or it can be viewed at URL
http://www.ukc.ac.uk/IMS/maths/people/P.M.Sutcliffe/preprints.html} 
 shows a surface of constant energy density
${\cal E}=0.06$ for the five values $a=-0.25,-0.18,0.00,+0.18,+0.25$.
We see that, indeed the energy density is initially localized in four regions
roughly centered on the vertices of a tetrahedron. Let us think of
these vertices as being opposite corners of a cube as in Fig 3.
On any one face of the cube the incoming energy density is concentrated
on two  opposite corners of the face (black spheres in
Fig 3.) 
and it flows around the edges of the
face until it is localized on the two remaining corners 
(white spheres in Fig 3.)  as the
monopoles separate. This suggests that a useful way to view this
scattering process is as pairs of $90^\circ$ scatterings occuring
simultaneously.

\section{Conclusion}
\news
\ \indent
Using a numerical scheme we have computed the energy densities
of a 3-monopole with tetrahedral symmetry and a cubic 4-monopole with
octahedral symmetry whose existence was recently proved \cite{HMM}.
We then proved the existence of a one parameter family of deformations
of the cubic 4-monopole which has tetrahedral symmetry. 
In the moduli space approximation this describes a 4-monopole
scattering process and we used our numerical scheme to analyse
this further.

There are a number of interesting aspects which remain in the
study of monopoles with the symmetries of regular solids.
One obvious task is to construct a family of 3-monopole solutions
which describes the scattering process in which the tetrahedral
3-monopole is formed. This problem is currently under investigation 
but is more difficult than the scattering
considered in this paper, since the family has only $C_3$ symmetry
which is not
as useful as the tetrahedral symmetry which allowed us to solve
the corresponding problem in the 4-monopole case. 
Another issue is that of a monopole with icosahedral symmetry.
A monopole with icosahedral symmetry has to have charge at least 
six, but in \cite{HMM} it was proved that no such monopole 
of charge six exists. We have proved that an icosahedral
monopole of charge seven exists and are currently investigating
its properties. These results and others on symmetric monopoles
will be presented elsewhere \cite{HSb}.\\

\noindent{\bf Acknowledgements}

Many thanks to Nigel Hitchin and Nick Manton for useful 
discussions. CJH thanks the EPSRC for a research studentship and the
British Council for a FCO award. PMS thanks the EPSRC
for a research fellowship.\\

\noindent{\sl Note added}.

 Recently the moduli space metric 
for the tetrahedrally symmetric 4-monopoles introduced in
this paper has been calculated, P.M. Sutcliffe,
 Phys. Lett. 357B, 335 (1995).

%\newpage

%\begin{figure}[ht]
%\begin{center}
%{\Large \bf }
%\vskip 1cm
%\leavevmode
%{\hbox
%1{\epsfxsize=5.8cm \epsffile{mon1fig4a.ps}}
%2{\epsfxsize=5.8cm \epsffile{mon1fig4b.ps}}
%}
%{\hbox
%3{\epsfxsize=5.8cm \epsffile{mon1fig4c.ps}}
%}
%\hbox{
%4{\epsfxsize=5.8cm \epsffile{mon1fig4d.ps}}
%5{\epsfxsize=5.8cm \epsffile{mon1fig4e.ps}}
%}
%\end{center}
%\caption{Four monopole scattering; surface of constant energy 
%density ${\cal E}=0.06$ for values
%(1) $a=-0.25$, (2) $a=-0.18$, (3) $a=0.00$, 
%(4) $a=0.18$, (5) $a=0.25$.}
%\end{figure}

%\newpage

\end{document}